# Are Gamma-Ray Bursts the Sources of the Ultra-High Energy Cosmic Rays?

Charles D. Dermer

*Code 7653, Naval Research Laboratory, 4555 Overlook Avenue, SW*
*Washington, DC 20375-5352 USA*

**Abstract.** A checklist of criteria is presented to help establish the sources of ultra-high energy cosmic rays (UHECRs). These criteria are applied to long-duration GRBs in order to determine if they are UHECR sources. The evidence seems to favor blazars and radio galaxies (or other sources) over GRBs.

## 1. INTRODUCTION

One of the outstanding unsolved problems in astronomy is the origin of the UHECRs. Gamma-ray bursts have been proposed as a likely source class [1]-[3]. In spite of major experimental advances that have occurred in the last decade, including the Pierre Auger Observatory, the IceCube South Pole Neutrino Detector, and the Swift and Fermi missions, progress towards solving this puzzle has been elusive. Yet progress has been made. Both the Auger [4] and HiRes [5] Observatories detect a spectral cutoff in the UHECR spectrum at particle energies $E \approx 5\times10^{19}$ eV, consistent with the GZK cutoff. This result favors an astrophysical explanation over a particle physics origin of UHECRs. The anisotropy in the arrival direction of UHECRs observed with Auger, though less significant than originally reported [6], is not aligned with the Galactic plane where Auger's exposure is greatest, but tends to align with the supergalactic plane. The strongest source is consistent with the location of Centaurus A [7] or the more distant Centaurus supercluster of galaxies [8], strengthening the likelihood of an extragalactic origin of the highest-energy cosmic rays.

Given that the sources of UHECRs are extragalactic (though this does not preclude a rare, unusual UHECR-forming event from taking place occasionally in the Milky Way), we can make a checklist of criteria that an UHECR source model should satisfy:

1. Extragalactic origin;
2. Mechanism to accelerate to ultra-high energies;
3. Adequate energy production rate per unit volume, or luminosity density;
4. Sources within the GZK radius; and
5. UHECR survival during acceleration, escape, and transport.

These criteria are applied to long-duration GRBs, and are indicated for other classes of GRBs or supernovae with relativistic outflows. The same checklist is considered for blazars and radio galaxies. I conclude that the data, though hardly unambiguous and not preventing both blazars and GRBs from accelerating UHECRs, favor an origin in the BL Lac and FR1 radio galaxies [9].

## 2. UHECRS FROM LONG DURATION GRBS

We now step through the checklist for long duration GRBs, keeping in mind that the same issues apply to blazars and radio galaxies.

## Extragalactic Origin

After the discoveries with BATSE and Beppo-SAX, no one doubts that GRBs detected by space-based observatories occur at extragalactic distances. To preserve the weak anisotropy in the arrival direction of the UHECRs measured with Auger requires that the intergalactic magnetic field (IGMF), to avoid scrambling the arrival direction and smearing the anisotropy, be smaller than some characteristic value $B_{IGMF}$. By imposing the condition that the deflection angle $\theta_{dfl} \approx d/(2r_L\sqrt{N_{inv}}) \ll 1$, where $d$ is the source distance, $r_L = E/QB_{IGMF}$ is the Larmor radius of a particle with $E = 10^{20} E_{20}$ eV is the particle energy and charge $Q = Ze$ is the particle charge, and $N_{inv}$ is the number of inversions of the magnetic field during transit from the source to Earth, one finds that

$$B_{IGMF}(nG) < \frac{3 E_{20}\sqrt{N_{inv}}}{Z(d/75\ Mpc)} \ . \tag{1}$$

For the specific case of Cen A at 3.5 Mpc,

$$B_{IGMF}(nG) < \frac{6(\theta_{dfl}/0.1) E_{20}\sqrt{N_{inv}}}{Z(d/3.5\ Mpc)} \ . \tag{2}$$

In either case, the limit $B_{IGMF} < 10\sqrt{N_{inv}}$ nG/Z are implied by the Auger arrival direction anisotropy, consistent with limits from Faraday rotation studies of radio quasars.

## Acceleration to Ultra-High Energies

Whether long duration GRBs or blazars can accelerate protons or ions to energies $\sim 10^{20}$ eV is another question. An elementary estimate based on the Hillas criterion—which is the minimal condition that a source must satisfy in order to be able to accelerate particles to ultra-high energies—shows that they can for general conditions governing Fermi acceleration processes. The Hillas criterion is that the Larmor radius must be smaller than the size scale of the system, which can be written as $r_L' = E'/QB' < R'$, where the primes refer to comoving frame quantities and $B'$ is the comoving magnetic field. After escaping from a source moving with Lorentz factor $\Gamma$, the maximum particle energy is $E_{max} = \Gamma E'_{max} < \Gamma E' = ZeB'R'\Gamma = ZeB'R$, noting that $R' = R/\Gamma$ from length contraction of the stationary frame size scale as measured in the comoving frame. Relating the magnetic field energy density by a factor $\varepsilon_B$ times the proper frame energy density associated with the wind luminosity $L$, then $B'^2 = 2\varepsilon_B L/(\beta c R^2 \Gamma^2)$, and

$$E_{max} < \left(\frac{Ze}{\Gamma}\right)\sqrt{\frac{2\varepsilon_B L}{\beta c}} \ , \tag{3}$$

implying

$$L_\gamma > \frac{3 \times 10^{45}}{Z^2} \frac{\Gamma^2}{\beta} E_{20}\ \mathrm{erg\ s^{-1}} \ , \tag{4}$$

noting that the apparent γ-ray luminosity $L_\gamma < L$. Thus UHECR sources should have $L_\gamma \gg 10^{45}/Z^2$ erg s$^{-1}$, and significantly more γ-ray luminosity if the sources have large $\Gamma$. This sketchy estimate can be improved by considering a specific model for particle acceleration. Within the context of colliding shells, which is a favored scenario for making the variable radiation in GRBs and blazars, equation (4) holds in case of a strong forward shock or relativistic forward and relativistic reverse shocks when the winds have large $\Gamma$ contrast, and are highly variable [9].

Gamma-ray observations provide a method, independent of superluminal motion observations and inferences from Compton-catastrophe arguments, to give the minimum bulk outflow Lorentz factor $\Gamma_{min}$. Many subtleties arise in the implementation of the γγ opacity test, e.g., the question of co-spatiality of the high energy γ ray with the target photons, the precise definition of variability timescale and its connection to the comoving size scale of the emitting region, the geometry of the emitting region, and the chance probability for detecting a high-energy photon from an absorbed region, but it is difficult to reduce estimates of $\Gamma_{min}$ obtained from naïve arguments applied to bright γ-ray sources with broadband coverage by more than a factor of $\sim 2 - 3$.

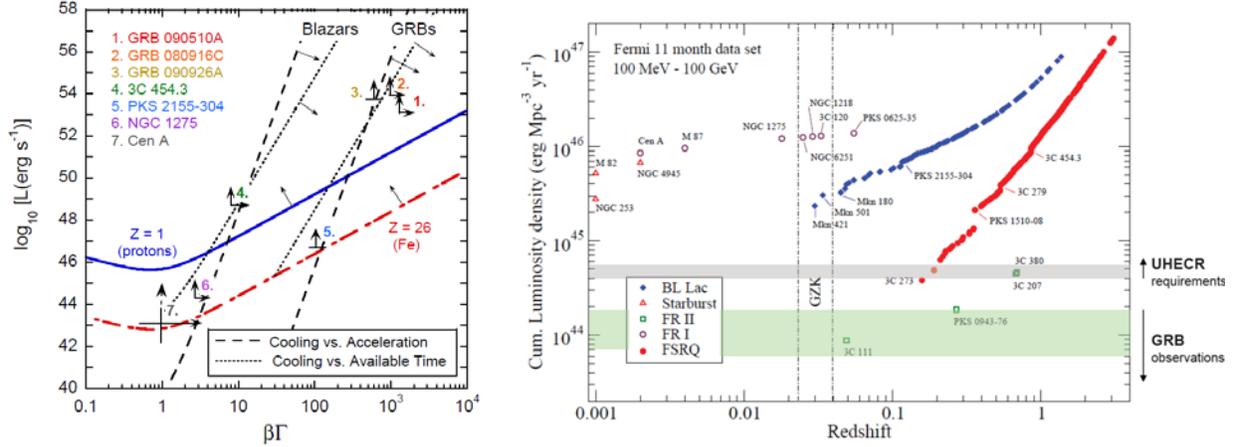

**FIGURE 1.** (a) Data shows apparent isotropic $L_\gamma$ versus $\beta\Gamma$ for blazars, radio galaxies, and GRBs. Solid and dot-dashed curves plot the constraint given by equation (2). (b) γ-ray luminosity density inferred from Fermi observations of various classes of γ-ray emitting AGNs, compared with the local luminosity of long-duration GRBs estimated here, and UHECR source requirements.

Figure 1(a) shows a plot of the source luminosity $L > L_\gamma$ against $\beta\Gamma = \sqrt{\Gamma^2 - 1}$ for blazars and GRBs. Note that GRB 090510A is a short, hard GRB, whereas the GRB 080916C and GRB 090926A are long-duration GRBs. The γ-ray luminosities of the sources are given at the time when $\Gamma_{min}$ was inferred. Except for PKS 2155-304, which uses HESS data for the giant outbursts in 2006, all the data were measured with the Large Area Detector on the Fermi Gamma-ray Space Telescope. As can be seen, the powerful Fermi GRBs have more than adequate luminosity to accelerate either protons or ions to ultra-high energies, even with Lorentz factors $\Gamma_{min} \sim 10^3$ as inferred from γγ opacity arguments. By comparison, the blazars and radio galaxies have smaller apparent luminosities and also smaller $\Gamma_{min}$. Acceleration of high-Z ions like Fe to ultra-high energies is possible for these sources on the basis of equation (4), but acceleration of protons appears unlikely, except possibly during flaring episodes.

## Luminosity Density

To maintain the measured flux of UHECR protons against photopion losses with the CMB requires a luminosity density in UHECRs of $\sim 10^{44}$ erg Mpc$^{-3}$ yr$^{-1}$ [10]. This value can easily be obtained by dividing the energy density of UHECRs at $\approx 10^{20}$ eV, which is $\approx 10^{-21}$ erg cm$^{-3}$, by the photopion loss timescale $t_{p\gamma} \approx 100$ Mpc/c. Because $t_{p\gamma}$ increases rapidly with decreasing energy, the required luminosity density remains at this level even when fitting to the ankle of the cosmic-ray spectrum at $E \approx 4\times 10^{18}$ eV as confirmed by direct modeling [11], though precise values of the local luminosity density depends in detail on the assumed cosmic star-formation rate factor applicable to GRBs.

The standard argument for inferring the luminosity density of a source class is to assume that the γ-ray luminosity represents, within some unknown factor, the luminosity in UHECRs. This seems reasonable because radiative losses into γ rays are likely to occur during acceleration. Moreover, the γ rays may underestimate the power in cosmic rays, because if the particles experience severe radiative losses during acceleration, then they could not be sources of the UHECRs [10]. This estimate is furthermore independent of beaming factor, because for every beamed source we detect, a proportional number of misdirected sources will point away from us. The argument suffers, however, from the likelihood that a large percentage of the γ radiation is produced by ultra-relativistic leptons (especially in the case of blazars where correlated variability with lower energy bands can be monitored). Moreover, it is not certain that MeV γ rays from long duration GRBs, which comprise the bulk of the energy output, is entirely nonthermal, inasmuch as a thermal/photospheric interpretation can potentially resolve the line-of-death problem that plagues nonthermal synchrotron interpretations of GRBs. Thus the nonthermal LAT emission might be a more appropriate luminosity with which to define the nonthermal luminosity density of GRBs [12].

The local luminosity density $\ell = 10^{44}\ell_{44}$ erg Mpc$^{-3}$ yr$^{-1}$ can be evaluated from the luminosity function [13,14] or from a physical model [15] for GRBs. Given the luminosity function $\Phi(L)$ and the mean duration $\Delta t$ of the GRB in the proper frame, then

$$\ell \approx \Delta t \int_0^\infty dL \, \Phi(L) \, . \tag{5}$$

Using this expression and assuming $\Delta t \approx 10$s, the treatment of Guetta et al. [13] gives $\ell_{44} \approx 0.05 - 0.08$ in the 50 – 300 keV band, which when corrected by a bolometric factor $\approx 6$ for the range 1 keV – 1 MeV, agrees with the value of $\ell_{44} \approx 0.3 - 0.4$ that Truong Le and I found in analysis of BATSE and Swift data [15]. It is also consistent with the treatment of Wanderman & Piran [14], who find $\ell_{44} \approx 0.6$. This value can be compared directly with the mean GRB flux $\phi$ in terms of the measured BATSE fluence per year, which is at the level of $\phi(>20 \text{ keV}) \approx 0.006$ erg cm$^{-2}$ yr$^{-1}$ [12]. A simple derivation shows that

$$\ell \approx k \frac{H_0}{c} \phi \approx 10^{43} \text{ erg Mpc}^{-3} \text{ yr}^{-1} \, , \tag{6}$$

where $k < 1$ is a cosmological factor dependent on the star-formation rate history of GRBs. The discrepancy between the two estimates may reflect the choice of $\Delta t$ relating the peak luminosity and the mean fluence. In either case, however, we derive a value for the local luminosity density of GRBs much less than the canonical value of $10^{44}$ erg Mpc$^{-3}$ yr$^{-1}$ required to power the UHECRs. There are ways to avoid this conclusion, which has been noted for a long time [16], for example, by assuming a large UHECR baryon load and low radiative efficiency, or the existence of missing untriggered GRBs that contribute to $\phi$ [17]. In any case, the nonthermal γ-ray luminosity of FR1 radio galaxies and BL Lac objects does not suffer from this problem, as seen in Fig. 1(b). Star-forming galaxies also have adequate nonthermal γ-ray luminosity related to cosmic-ray production, but they are individually far too weak, $<10^{40}$ erg s$^{-1}$, to be able to accelerate UHECRs according to equation (4).

## Sources within the GZK Radius

There will, of course, be GRBs that take place inside the GZK radius, taken here to be ~100 Mpc (a strong magnetic field will shrink the effective radius [18], though it cannot be so strong that the UHECR arrival directions are completely isotropized). But the rate of long-duration GRBs taking place in any given galaxy is very low, on the order of one event every $10^5$ year in an L$_*$ galaxy like the Milky Way. Because GRBs are impulsive rather than persistent events, the acceleration of UHECRs takes place quickly, and moderately strong magnetic fields are required to delay the arrival times of the UHECRs. These magnetic fields can be in intergalactic space or in the filamentary structure formation regions.

The local beaming corrected rate density of long-duration GRBs is $\approx 10$ Gpc$^{-3}$ yr$^{-1}$ [15], roughly consistent with the product of the apparent rate density, $\approx 0.5$ Gpc$^{-3}$ yr$^{-1}$ and the beaming factor of $\approx 75$ [13]. The GRB rate within the GZK radius is therefore $\approx 0.1$ yr$^{-1}$, of which $\approx 10^{-3} \theta_{-1}^2$ yr$^{-1}$ will have a jet with opening angle of $0.1\theta_{-1}$ rad pointing in our direction. If one naively assumes that the particles travel only through the intergalactic space, then their arrival times are extended by

$$\Delta t \approx \frac{d}{6c} \theta_{dfl}^2 \approx 400 \frac{Z^2 B_{-15}^2 d_{100}^3}{E_{20}^2 N_{inv}} \text{ s} \, , \tag{7}$$

where d = 100 d$_{100}$ is the distance to the source and $B_{IGMF} = 10^{-15}$ B$_{-15}$ G. Moreover, the deflection angle is only $\theta_{dfl} \approx 10^{-6}$ ZB$_{-15}$d$_{100}$/E$_{20}$, using our earlier expression. Thus it would seem that for IGMFs weaker than $B_{IGMF} \approx 10^{-15}$ G, UHECRs would come in short bursts and point back directly to their sources. To extend their arrival times over ~$10^5$ yr would then require significantly stronger IGMFs with $B_{IGMF} \sim 10^{-10}$ G/Z. If the UHECRs travel through regions associated with filamentary structure that contain fields reaching to $\approx 0.1$ μG on Mpc scales, however, then the arrival direction can be delayed even if the magnetic fields in intergalactic space are very weak [19,20].

## UHECR Survival and Escape from the Acceleration Site

A final requirement for an UHECR source is that the cosmic-ray particles can be accelerated and escape from the acceleration site without losing energy through radiative or adiabatic losses, in the case of UHECR protons, or through photodisintegration breakup in the case of ions. If UHECRs are proton-dominated, as claimed by HiRes, then a convenient method for escape is through neutron intermediaries formed in photopion processes that subsequently decay into protons far from the acceleration site, with an accompanying flux of neutrinos and cascading γ rays [21]. Ions will survive acceleration to in the external shocks, and depending on specific conditions, in the internal shocks of GRBs and the sub-energetic hypernovae [22]. Whether charged particles from an impulsive GRB source can escape from the acceleration site without in turn generating a cosmic-ray induced shock or losing energy through streaming instabilities has not been convincingly established.

## 3. OTHER SOURCE CLASSES

This checklist can be applied to other source classes, including the short hard GRB class, the X-ray flashes, the sub-energetic GRBs, and now the engine-driven supernovae with relativistic outflows that lack GRB-type emissions [23]. These classes, though they have much larger local rate densities than long-duration GRBs [20], are not much preferred over long-duration GRBs on the basis of their local luminosity density. Low-luminosity, sub-energetic GRBs, for example, have a large luminosity density in ejecta kinetic energy, estimated to exceed $10^{46}$ erg Mpc$^{-3}$ yr$^{-1}$ [24], but smaller or comparable nonthermal γ-ray luminosity densities than long-duration GRBs [25]. The same applies to the short hard class of GRBs, or to the nonthermal luminosity density inferred from the radio emissions of engine-driven supernovae.

Does this mean that blazars or other source classes are then favored? Though there are FR1 radio galaxies within the GZK radius, and the nonthermal luminosity density is 1 – 2 orders of magnitude in excess of what is needed to power the UHECRs, as seen in Fig. 1(b), the evidence is still not compelling. Much of the γ-ray luminosity in TeV blazars and radio galaxies is well-explained by a standard leptonic synchrotron/SSC model, leaving only a residual fraction to originate from hadronic processes. Nor do the arrival directions of UHECRs clearly single out radio galaxies or BL Lac objects except in the case of Centaurus A; in fact, correlation of arrival directions with Fermi 1LAC sources points to a range of galaxy types, including the starburst galaxy NGC 253 and the starburst/Seyfert galaxy NGC 4945 [26].

## 4. SUMMARY

Missing from this checklist are two additional items that might provide the most compelling arguments that GRBs are the sources of UHECRs. The first is the presence of electromagnetic signatures of UHECRs in their spectra. I recently reviewed this issue in light of the Fermi observations [11], and concluded that the evidence for ultra-relativistic hadrons from GRB γ-ray data is ambiguous, if not in favor of leptonic models. High-energy neutrino detections from GRBs from IceCube would be most compelling, even granted that if PeV neutrinos were detected, this would reveal the existence of ~$10^{16}$ – $10^{18}$ eV cosmic rays rather than UHECRs. The latest report from IceCube gives no grounds for optimism [27], nor do the large Lorentz factors inferred from Fermi observations of GRBs, which imply dilute internal target photon fields and low neutrino production efficiency. Yet GRB 090926A does show a power-law break, leading to an estimate of $\Gamma \sim 200 - 700$ [28] if the break is due to γγ effects. The existence of GRBs with bright GBM emission and suppressed LAT flux [29] points to the possibility of low-$\Gamma$ factor GRBs with large neutrino efficiency. One such extraordinary event, which might occur tomorrow, could transform our thinking. This is not mere wishful thinking—GRB science has seen many such events.

## ACKNOWLEDGMENTS

I thank Sayan Chakraborti, David Eichler, Kohta Murase, and David Wanderman for interesting discussions and comments, and Soebur Razzaque for collaboration. The work of C.D.D. is supported by the Office of Naval Research.

## REFERENCES

1. E. Waxman, *Physical Review Letters,* **75**, 386 (1995)
2. M. Vietri, *Astrophysical Journal*, **453**, 883 (1995)
3. C. D. Dermer, *Astrophysical Journal*, **574**, 65 (2002)
4. R. U. Abbasi et al., *Physical Review Letters*, **100**, 101101 (2008)
5. J. Abraham, et al. *Physical Review Letters*, **101**, 061101 (2008)
6. The Pierre Auger Collaboration, *Science*, **318**, 938 (2007)
7. P. Abreu, et al., *Astroparticle Physics*, **34** 314 (2010)
8. G. Ghisellini, G. Ghirlanda, F. Tavecchio, F. Fraternali, and G. Pareschi, *MNRAS*, **390**, L88 (2008)

9. C. D. Dermer and S. Razzaque, *Astrophysical Journal*, **724**, 1366 (2010)
10. E. Waxman and J. N. Bahcall, *Physical Review D,* **59**, 023002 (1999)
11. C. D. Dermer, in *Deciphering the Ancient Universe with Gamma-Ray Bursts*, edited by N. Kawai and S. Nagataki, New York: AIP, 2010, pp. 191-199
12. D. Eichler, D. Guetta, and M. Pohl, *Astrophysical Journal*, **722**, 543 (2010)
13. D. Guetta, T. Piran, and E. Waxman, *Astrophysical Journal*, **619**, 412 (2005)
14. D. Wanderman, and T. Piran, *Monthly Notices of the Royal Astronomical Society*, **406**, 1944 (2010)
15. T. Le, and C. D. Dermer, *Astrophysical Journal*, **661**, 394 (2007)
16. F. W. Stecker, *Astroparticle Physics*, **14**, 207 (2000)
17. E. Waxman, arXiv: 1010.5007 (2010)
18. T. Piran, arXiv : 1005.3311 (2010)
19. H. Takami, H. Yoshiguchi, and K. Sato, *Astrophysical Journal*, **639**, 803 (2006)
20. K. Murase and H. Takami, *Astrophysical Journal*, **690**, L14 (2009)
21. A. M. Atoyan and C. D. Dermer, *Astrophysical Journal*, **586**, 79 (2003)
22. X.-Y. Wang, S. Razzaque, and P. Mészáros, *Astrophysical Journal*, **677**, 432 (2008)
23. S. Chakraborti, A. Ray, A. Soderberg, A. Loeb, and P. Chandra, *Nature Communications*, arXiv:1012.0850 (2010)
24. X.-Y. Wang, S. Razzaque, P. Mészáros, and Z.G. Dai, *Physical Review D*, **76**, 083009 (2007)
25. K. Murase, K. Ioka, S. Nagatak, and T. Nakamura, *Physical Review D*, 78, 023005 (2008)
26. R. S. Nemmen, C. Bonatto, and T. Storchi-Bergmann, *Astrophysical Journal*, **722**, 281 (2010)
27. IceCube Collaboration, et al., arXiv:1101.1448 (2011)
28. A. Abdo, et al., *Astrophysical Journal*, in press, arXiv :1101.2082 (2011)
29. D. Guetta, E. Pian, and E. Waxman, *Astronomy & Astrophysics*, **525**, 53 (2011)